\documentstyle[12pt,psfig]{article}
\begin{document}
\begin{center}
\textbf{{\Large{}\textbf {Effects of temperature and sulphur on the
composition profile of Pt-Rh  nanocatalysts : A comparative study.}}}
\end{center}
\begin{center}
\textbf{Abir De Sarkar and Badal C. Khanra}\textbf{$^{\rm{}*}$}
\end{center}
\begin{center}
\textit{Condensed Matter Physics Group, Saha Institute of Nuclear Physics,}
\end{center}
\begin{center}
\textit{ 1/AF, Bidhannagar, Calcutta - 700064, India}
\end{center}

\noindent
\textbf{Abstract}

\noindent         
     Monte-Carlo simulation technique has been used to investigate the
     effect of temperature and adsorbed gases on the composition
     profile of  unsupported Pt-Rh nanocatalysts. For a 2406 atom fcc
     cubo-octahedral  Pt$_{\rm{}50}$Rh$_{\rm{}50}$ nanocatalyst  the
     shell-wise composition for all the eight shells has been
     simulated.  For the temperatures 700 K, 1000 K and 1300 K, the
     top shell of  clean  Pt-Rh  nanocatalysts  is found to  be
     mildly  Pt-enriched, while the second shell is Pt-depleted. The
     Pt concentration of the top shell  shows a  maximum at T = 1000
     K.  In presence of  a quarter monolayer  of adsorbed oxygen   the
     top shell shows Rh enrichment, while all the other shells show
     Pt-enrichment. This is true for all the three temperatures for
     which the composition profiles have been studied.

\noindent
PACS: 61.46.+W, 61.66Dk,68.10. Jy, 68.35. Md

\noindent
\textit{Keywords:}  Nanocatalysts, segregation, Monte-Carlo
simulation, adsorption

\noindent

$\bullet$ Corresponding Author\\
       FAX (91) (33) 337 4637\\
       Email: <badal@cmp.saha.ernet.in>

\newpage
\noindent
\textbf{1. Introduction}

\noindent
The ceria-supported Pt-Rh bimetallic catalysts have been found to be
very effective three-way catalysts for  simultaneous elimination of
the CO , NO and the hydrocarbons  from automobile exhaust gases [1-3].
These catalysts are increasingly used globally in the catalytic
converters of automobiles. It is important, therefore, to have a
thorough understanding of the structural, electronic and catalytic
properties of these catalysts. In this work , we present the results
of our theoretical investigations  on the composition profile of
unsupported Pt-Rh  bimetallic nanostructures  as a function of
temperature and adsorbates.

            The structural and catalytic properties  of the Pt-Rh
            alloy  surfaces have earlier been exhaustively studied  by
            several groups [4-11].  Van Delft et al. [4] used AES to
            investigate the Pt$_{\rm{}62 }$Rh$_{\rm{}38 }$   and
            Pt$_{\rm{}10 }$Rh$_{\rm{}90}$    alloy systems. They found
            a surface enrichment with Pt for both the systems. When
            studied as a function of temperature this surface
            enrichment showed a maximum at about 1000 K for both the
            systems. This phenomenon  was explained in terms of a
            vibration entropy contribution to the enthalpy.  Tsong et
            al. [5] made a layer-by-layer analysis of a Pt-Rh alloy
            tip at 1000 K to show that the topmost layer is
            Pt-enriched and the second layer is Pt-depleted.  The
            distribution of Pt and Rh atoms  depend on the annealing
            temperature.  Hirano et al. [6] , on the other hand,
            showed that  the  presence of adsorbates like  oxygen
            and/or nitric oxide  has a more pronounced effect on the
            chemical restructuring process  of the Pt-Rh surface
            layers  than the annealing temperature.   While there have
            been  a large number of theoretical studies on this
            segregation phenomenon in clean and gas-covered alloy
            surfaces [4-7], the number of segregation studies for
            supported particles are comparatively few. Some
            theoretical investigations for  Pt-Rh  particles have been
            made by  Strohl and King [8], Schoeb et al. [9] and Yang
            et al [10].  These works are based either on rigid-lattice
            Monte-Carlo (MC)  simulations or  on molecular dynamics /
            Monte-Carlo-correlated effective medium theory, where the
            interaction energies were  generated from a simplified
            version of the non-selfconsistent
            electron-density-functional correlated effective medium
            (CEM) theory [11]. For  the fcc cubo-octahedral particles
            the  atomic distribution at the corner, edge and fcc(100)
            and (111)-like faces were obtained as a function of  the
            particle size and composition.  A  shell-by-shell
            composition profile  of the particles were  not, however,
            studied in these works.  It is the main purpose of the
            present work to  investigate  the shell-wise composition
            of   Pt-Rh  nanoparticles  as a function of temperature
            and  adsorbate coverage. For  the 2406 fcc cubo-octahedral
            Pt-Rh  nanoparticles  a Monte-Carlo simulation  has been
            carried out  to analyse the  atomic distribution in top
            few layers . We show that the presence of  adsorbate like
            oxygen  drastically changes the composition profile, while
            the effect of temperature from 700K-1300 K  on the
            composition profile is relatively much small.  These
            studies  are useful  in understanding the  structure and
            catalytic properties of  ceria-supported  Pt-Rh catalysts
            used in the catalytic converter . The plan of the paper is
            as follows.  In section 2,  the  theoretical  model
            developed earlier and used in the present  work  is
            briefly outlined. The results for clean and
            oxygen-adsorbed Pt-Rh nanoparticles are presented in
            section 3.  The conclusions are drawn in section 4.

\noindent
\textbf{2. Theoretical model}

\noindent
 The theoretical model used in this work is the MC simulation
 technique employed elsewhere in connection with surface  segregation
 in  presence and in absence  of adsorbed species[12-16].   One
 determines the pair-bond energy between two nearest neighbour atoms
 E$_{ij}$   in the form

\begin{equation}
E_{\rm{}ij}  = w_{\rm{}ij  }/Z  + E_{\rm{}c}^{\rm{}i} (n)/n  + E_{\rm{}c}^{\rm{}j}  (m) /m
\end{equation}                                          

\noindent
where  the indices i  and   j are the component metals A or B of the
bimetallic system AB ; and w$_{ij}$   is the interchange  energy  for
two dissimilar atoms. w$_{ij}$ = 0  if i = j  and this interchange
energy is related to the  excess heat of mixing.  Z is the  bulk
coordination number. E$_{c}^{i}$ (n) is the partial bond energy of the
i-atom  with n nearest neighbours etc.   Usually, the
coordination-dependent partial bond energy , E$_{\rm{}c}$$^{\rm{}i}$
(n) ( for  n = 2 - 12),  is calculated from an  empirical  formula
like

\begin{equation}
E_{\rm{}c}^{\rm{}i} (n) = a^{\rm{}i}  + b^{\rm{}i} n + c^{\rm{}i} n^{\rm{}2}   
\end{equation}

\noindent
where a$^{\rm{}i}$, b$^{\rm{}i}$  and c$^{\rm{}i}$  are adjustable
constants that can be derived from  experimental   data of dimer
energy ,   surface energy and the cohesive  energy of a metal
corresponding to  n=2, 8 and 12   respectively.   Once the pair
binding energy E$_{\rm{}ij}$  is  evaluated it  is the usual MC
procedure to  compute the  configuration energy for a particular
distribution of  A and B atoms in the nanoparticle.  And from various
such distributions   an equilibrium distribution may be  found giving
the shell-wise occupation of sites by A or B atoms. In presence of
adsorbates, there are additional  energy terms to describe the
chemisorptive bond and the concomitant  metal-metal  pair-bond energy
variations.  King and Donnelly [12]  showed that since in Monte-Carlo
simulations only the differences in energies are calculated, it is
sufficient to add the difference in chemisorption energy of the
adsorbate on the two metals  to the configuration energy.  This has
been done in this work to study the role of adsorbed oxygen on the
composition profile.

\noindent
\textbf{3. Results and Discussions}

\noindent
\textbf{   }We are interested in the composition profile of the outer
shells of the Pt-Rh nanoparticles  since it is only the outer shells
which  are important  in catalysis and can be influenced  by adsorbed
atoms.  For  convenience, we  avoid the usual description  of shell
structure  where the shells are constructed by atoms equidistant  from
the centre. As may be found from the work of Yang and DePristo [10]
that  to describe a  200-atom fcc cubo-octahedral particle one needs
11 shells.  This concept  may be very useful  for describing the
stability and morphological structure of a nanoparticle. However,  for
a large fcc cubo-octahedron particle of 2406 atoms as considered in
this work this  description would require a large number  of shells
which is very difficult to handle.   In our treatment , the outermost
shell (we also call it `top shell') comprises  all the atoms at the
corner sites, bridge sites, fcc(100) and fcc(111)-type faces. The
total number of such atoms is 752. If we peel off these atoms  from
the outermost shell, the second shell would be exposed to the adsorbed
gases. The number of atoms in the second shell is 582. In this way, we
may have just eight shells  with total number of atoms in the shells (
counting from the outermost shell inward) as 752,  582, 432,  302,
192,   102,   38  and  6 respectively.  The number of Pt and Rh atoms
are counted in each shell and the  fraction of these atoms in the
shells are calculated.  In this work we have calculated the fraction
of Pt and Rh atoms for a bulk Pt$_{\rm{}50}$Rh$_{\rm{}50}$  particle.

\noindent
The values of the parameters used  for the present  MC simulation are
the following:

   Pt:  a$^{\rm{}   }$ = -0.4287398,    b$^{\rm{} }$ = -0.0474969,
   c$^{\rm{} }$ =  0.0035483 , E$_{\rm{}ad }$(O) = 85 kcals/mole

  Rh:  a$^{\rm{}   }$ = -0.4744544,   b$^{\rm{} }$ = - .0341727,
  c$^{\rm{}  }$ =  0.0028138 , E$_{\rm{}ad }$(O)  = 102 kcals/mole

\noindent
  and w$_{\rm{}ij}$ /Z = -0.00691 eV.    E$_{\rm{}ad}$(O) is the
  chemisorption  energy of oxygen  atom on the metal.

\noindent
We have  calculated the composition profile for  T = 700 K, 1000 K and
1300 K.  The studies at  T = 700 K  is  of particular  importance
since the exhaust  temperature at the three-way catalyst is of this
order.  To study the effect of  adsorbed gases we have considered
one coverage (  = 0.25 )  of adsorbed oxygen. This is because it is
the presence or absence of oxygen that  controls the CO oxidation and
NO reduction in three-way catalysts. The calculation , however, can
easily be repeated for other coverages and other adsorbed gases . The
present work attempts to study  the qualitative change  of composition
profile as a function of temperature and  adsorbate coverage.

\noindent
         Figure 1  shows the \% of Pt atoms in top four shells.  The
         top shell ( shell number 1) corresponds to the surface layer
         and shows  Pt-enrichment for all the three temperatures;
         while the second and third shell shows Pt-depletion for the
         three temperatures.  The no segregation line is represented
         by the dots.  The behaviour from shell number 4 to higher (
         moving towards the core of the particle ) is
         divergent. However,  for surface and catalytic properties
         the top 2-3  shells are of real interest since they
         constitute the surface layers . Besides, the top shell for a
         2406-atom particle represents 31\% of total atoms, sometimes
         known as dispersion; while the top 2 , top 3  and top 4
         shells  represent 55\% , 73\% and  86\% of total atoms for
         this particle respectively. If we  compute the   \%  of Pt
         atoms averaged over top 2, top 3 and top 4 shells,  the
         results would be as shown in Figure 2. The results are
         computed for all the three temperatures. The results averaged
         over all the eight shells would ,of course,  be the bulk Pt
         concentration which is 50\%. This is why all the curves meet
         at  shell number 8. It may be noted that  for T = 700 K the
         top shell is Pt-enriched; but  the averages over top 2, top 3
         and top 4 shells show Pt-depletion.  For T = 1000 K and 1300
         K, however, the averages over top 2, top 3 and top 4 shells
         show Pt-enrichment.  This signifies  that the properties,
         that depend on the composition of the top shell only,  would
         not change with temperature, since the shell is
         Pt-enriched. However, the properties that depend on the
         average composition of top 2 , top 4 or top 4  shells will
         change with temperature.  In Figure 3 we  plot the  same
         shell-averaged Pt concentration as a function of
         temperature. In  addition to the features described  above
         for  Figure 2, an  important  feature becomes obvious in
         figure 3. It is that  the Pt-enrichment shows a maximum  at T
         = 1000 K.  This  result agrees very well with the
         experimental findings of  Van delft et al. [4] for the
         polycrystalline alloy surface. The authors [4] explained this
         behaviour with a surface segregation model containing both
         enthalpy and vibrational entropy contribution.  It is likely
         that  for  the  present system of nano-sized Pt-Rh  particles
         also the entropy contribution may be responsible for the
         temperature dependence of  surface segregation.   In Figure 4
         we plot the \% of Pt atoms in presence of 0.25 monolayer of
         adsorbed oxygen and compare them with the results for clean
         system, shown already in Figure 1. It may be noticed that the
         presence of 0.25 monolayer of oxygen has a very strong effect
         on the  composition of the top shell as well as on other
         shells.  Because of higher binding  energy  of oxygen on Rh
         compared to that on Pt the Rh atoms are pulled to the top
         shell at all the three temperatures. This segregation is so
         strong that  the  rest of the shells of the particle become
         Pt-enriched.   On comparing the effect of temperature and
         adsorbates  on the shell composition it may be noted from
         Figure  4  that the effect of temperature  on the composition
         profile  is mild. This is obvious from the spreadth of the
         composition for three temperatures in both the clean and
         adsorbed cases.   However, the  large separation of the two
         sets of curves ( each set consisting of the results for three
         temperature), one for the clean case and the other for the
         oxygen-adsorbed case,   signifies  that the effect  of
         adsorbates is much stronger on the composition.  We believe,
         these results can be exploited  in the three-way catalyst
         where Rh acts as the main component in reducing NO.
         Furthermore, it may be mentioned that the present study
         considers one bulk composition and one coverage of one
         adsorbed species. This may be repeated for other
         compositions, adsorbates and coverages as the situation
         demands.

\noindent
\textbf{4. Conclusions}

\noindent
   A Monte-Carlo procedure has been  utilised  to compute the  atomic
   composition of a Pt-Rh nanoparticle  in  the form of shells
   gradually peeled off from the outer layer. The composition profile
   correctly predicts Pt segregation in the top layer ; but the
   variation of the composition of the outer shells with temperature
   is small.  In contrast, the variation of  the composition of the
   shells with presence of  adsorbed oxygen  is very large.

\newpage
\noindent
\textbf{References}

\noindent
[1]   K. C. Taylor, \textit{In}  Automotive Catalytic Converters,
Springer, Berlin(1984).

\noindent
[2]   K. Taschner, \textit{In}  Catalysis and Automotive Pollution
Control II, Ed. A. Crucq, Elsevier, Amsterdam,(1991),p-17.

\noindent
[3]   C. Howitt, V. Pitchon, F. Garin and G. Maire, \textit{In}
Catalysis and Automotive Pollution  Control III, Eds. A. Frennet, and
J. M. Bastin, Elsevier, Amsterdam (1995), p-149.

\noindent
[4] F. C. M. J. M. Van Delft, Surf. Sci. 189/190 (1987) 1129.

\noindent
[5] T. T. Tsong, D. M. Ren and M. Ahmad, Phys. Rev. B38 (1988)7428.

\noindent
[6] H. Hirano, T. Yamada, K. Tanaka, J. Sierra and B. E. Nieuwenhuys,
Vacuum 41(1990) 134;  H Hirano, T. Yamada, K. Tanaka, J. Sierra and
E. Nieuwenhuys, Surf. Sci. 222(1989)L804; H Hirano, T. Yamada,
K. Tanaka, J.  Sierra and  B. E. Nieuwenhuys, Surf. Sci. 226(1990)1.

\noindent
[7] F. C. M. J. M. Van Delft, B. E. Nieuwenhuys, J. Sierra and
R. M. Wolf, ISIJ International, 29(1989)550.

\noindent
[8] J. K. Strohl and T. S. King, J. Catal. 116 (1989)540,
\textit{ibid}. 118(1989)53.

\noindent
[9] A. M. Schoeb, T. H. Raeker, L. Yang, X. Wu,  T. S. King and
A. E. DePristo, Surf. Sci. Lett 278(1992) L125.

\noindent
[10]    L. Yang and A. E. DePristo, J. Catal. 148 (1994)575.

\noindent
[11] M. S. Stave, D. E. Sanders, , T. J. Raeker and A. E. DePristo
J. Phys. Chem. 93(1989) 1556.

\noindent
[12]    T. S. King and R. G. Donnely, Surf. Sci.141(1984)417.

\noindent
[13]    B. C. Khanra and M. Menon, Chem. Phys. Lett. 305(1999)89.

\noindent
[14]    B. C. Khanra and  M. Menon, Physica B 291(2000)368.

\noindent
[15] A. De Sarkar, M. Menon and B. C. Khanra, Appl. Surf. Sci. (In press)        

\newpage

\noindent
\textbf{{\large{}\textbf{Figure Caption}}}
\vskip 1.2cm
\noindent
\textbf{{\large{}Fig. 1}}{\large{}.  MC-simulated  \% of Pt atoms in
different shells of a 2406-atom
Pt}$_{\rm{}{\large{}50}}${\large{}Rh}$_{\rm{}{\large{}50}}${\large{}
nanoparticle  for different temperatures .}

\noindent
\textbf{{\large{}Fig. 2}}{\large{}. MC-simulated average  \% of Pt
atoms in top, top 2, top 3 and top 4    shells of a 2406-atom
Pt}$_{\rm{}{\large{}50}}${\large{}Rh}$_{\rm{}{\large{}50}}${\large{}
nanoparticle  for different temperatures.}

\noindent
\textbf{{\large{}Fig.3}}{\large{}. MC-simulated average  \% of Pt
atoms in top, top 2, top 3 and top 4   shells of a 2406-atom
Pt}$_{\rm{}{\large{}50}}${\large{}Rh}$_{\rm{}{\large{}50}}${\large{}
nanoparticle  plotted as a function of  temperature to demonstrate the
existence of a maximum at T = 1000 K.}

\noindent
\textbf{{\large{}Fig. 4.}}{\large{} MC-simulated  \% of Pt atoms in
different shells of a clean  and oxygen-adsorbed 2406-atom
Pt}$_{\rm{}{\large{}50}}${\large{}Rh}$_{\rm{}{\large{}50}}${\large{}
nanoparticle  . }

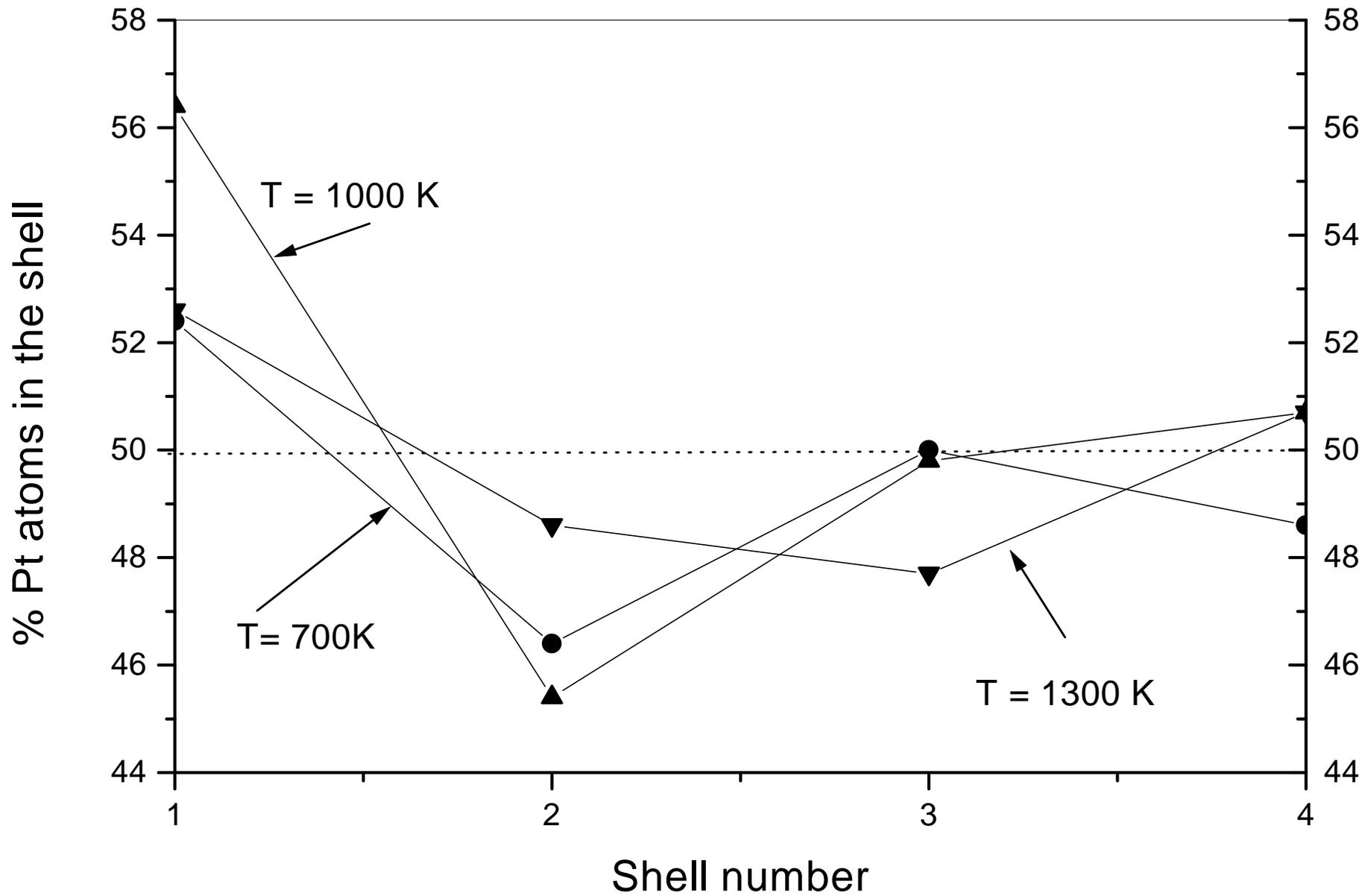
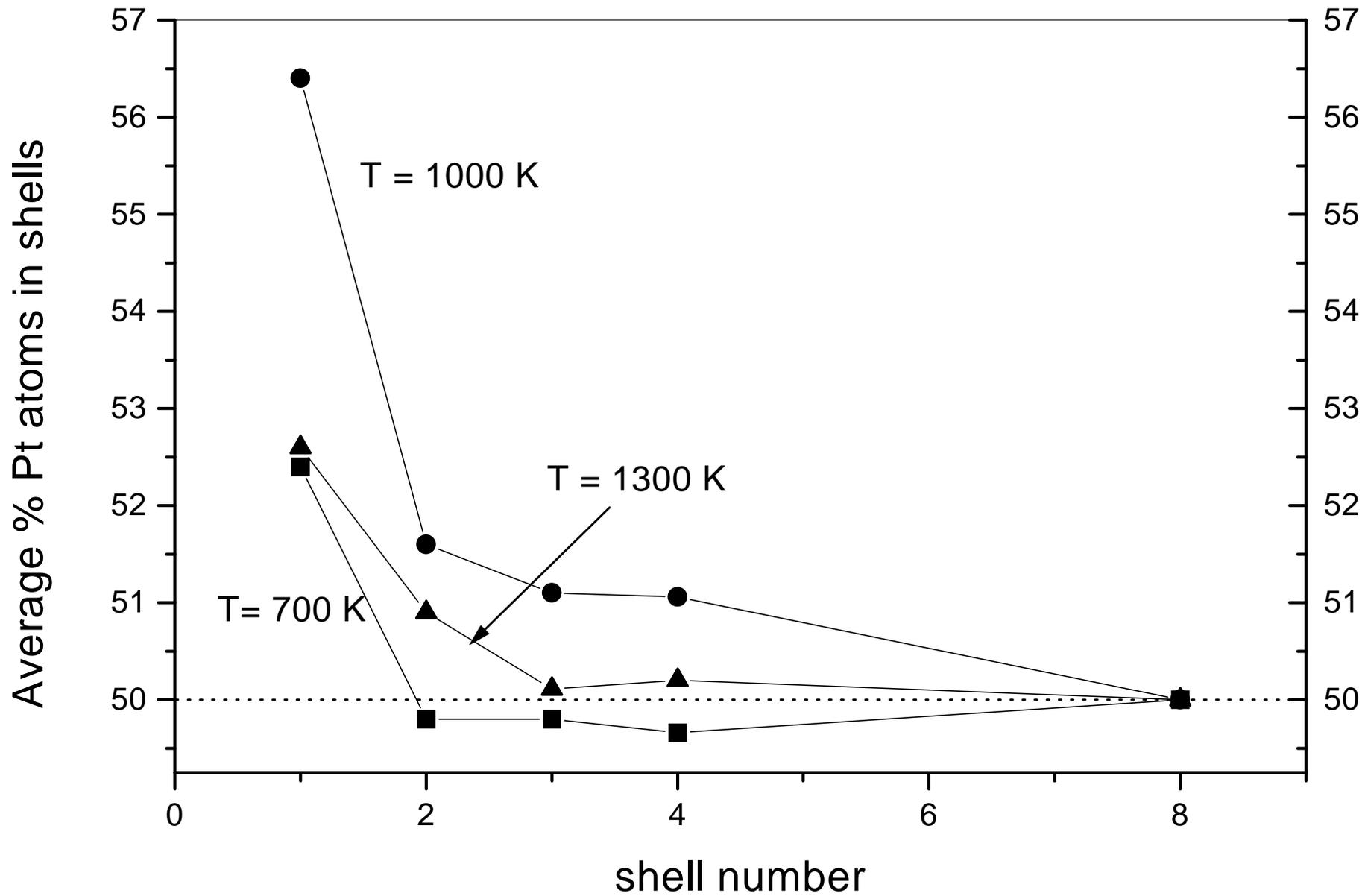
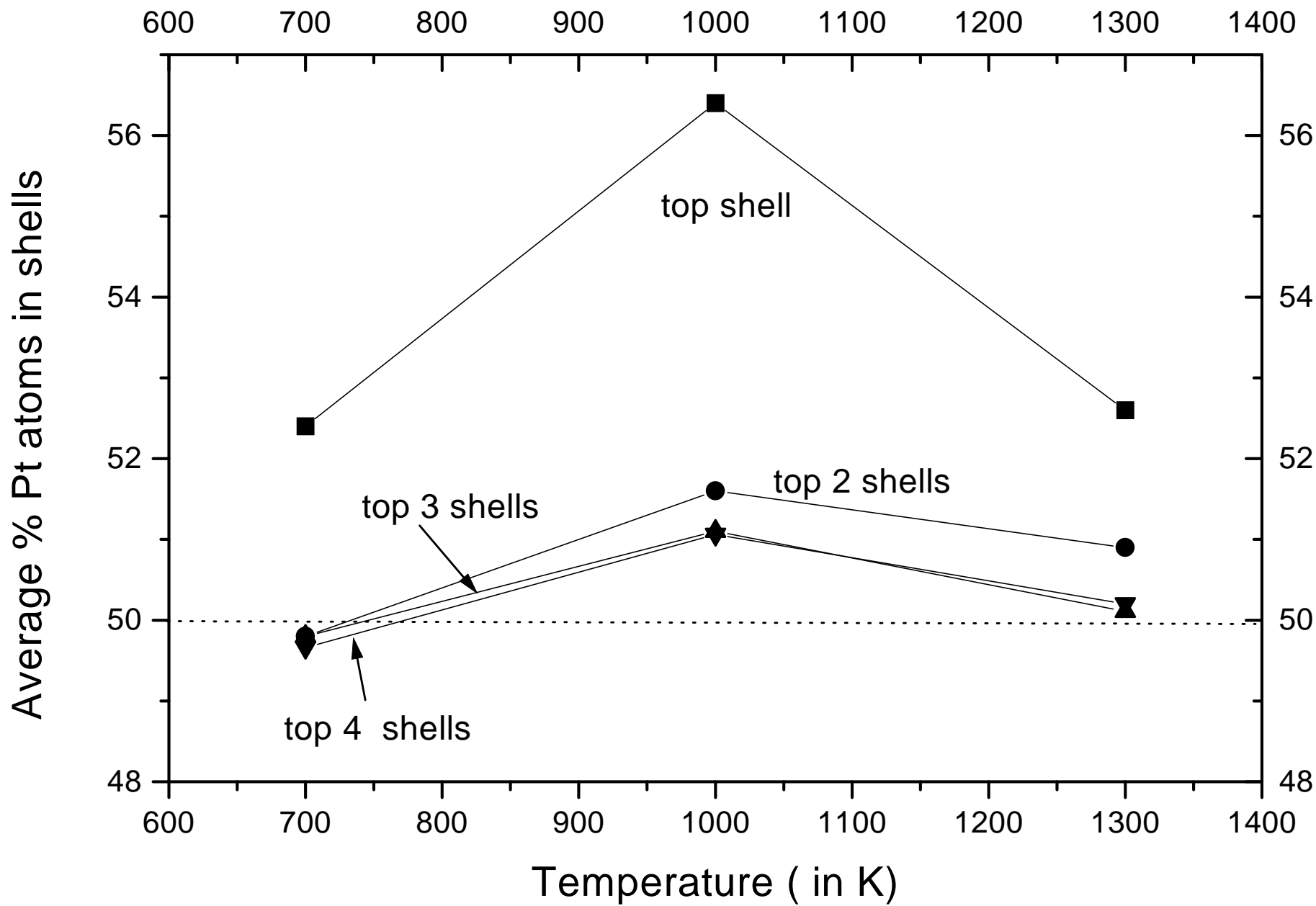
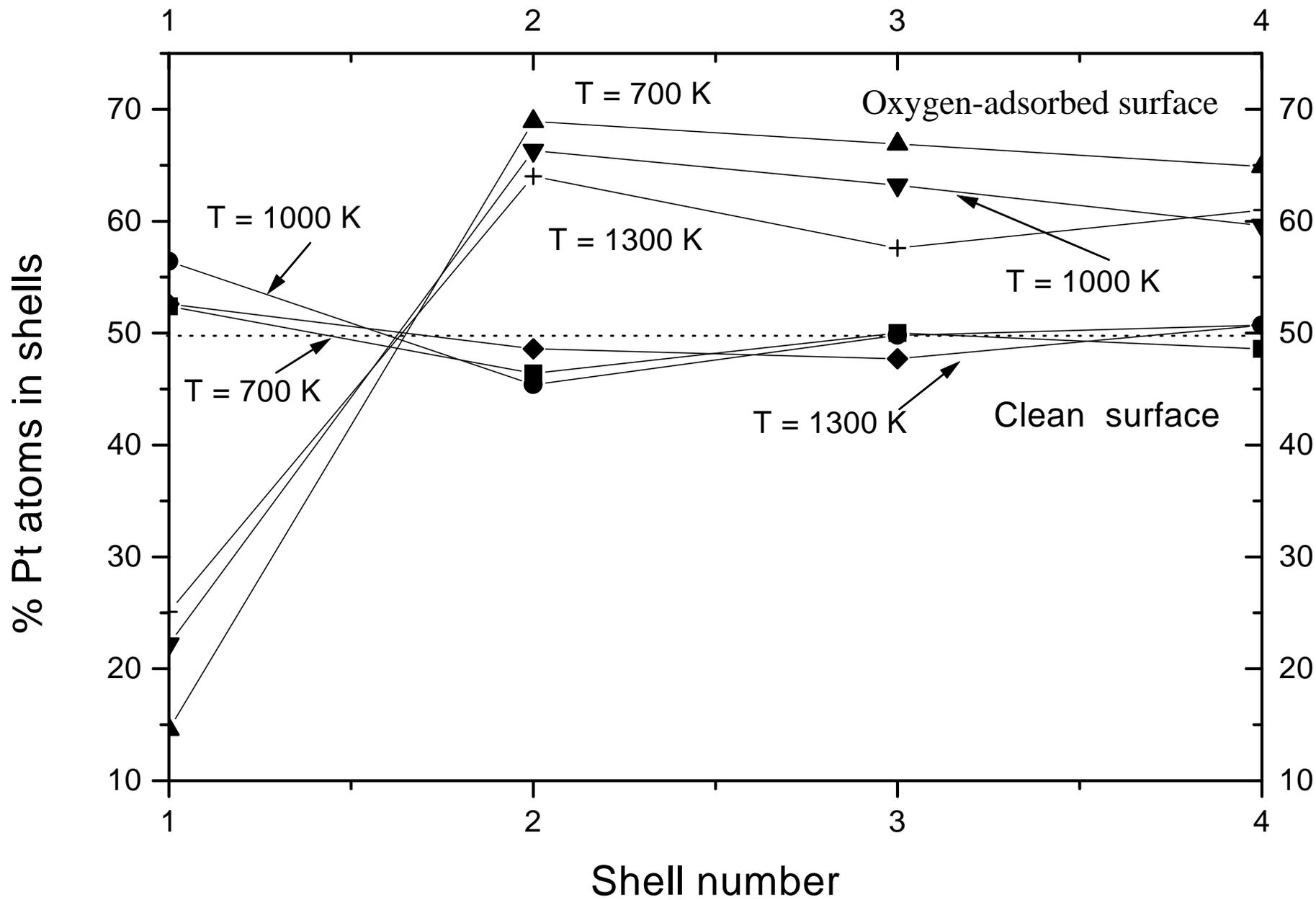
\end{document}